\documentclass[prd,aps]{revtex4}
\usepackage[utf8]{inputenc}
\usepackage{amssymb}
\usepackage{amsmath}
\usepackage{cancel}
\usepackage{mathtools}
\usepackage{graphicx}
\usepackage{epsfig}
\usepackage{bm}
\usepackage{amsfonts}
\usepackage{color}
\usepackage{subfigure}
\usepackage{epsf}
\usepackage{epsfig,graphicx}
\usepackage{url}

\begin{document}
\title{ScalPy:  A Python Package For Late Time Scalar Field Cosmology}

\author{Sumit Kumar}
\email{sumit@ctp-jamia.res.in}
\affiliation{Centre for Theoretical Physics\\Jamia Millia Islamia, New Delhi 110025, India}

\author{Abhishek Jana}
\email{abhishek11001@iiserkol.ac.in}
\affiliation{Department of Physical Sciences\\ Indian Institute of Science Education And Research Kolkata\\ Mohanpur, West Bengal 741246, India}

\author{Anjan A. Sen}
\email{aasen@jmi.ac.in}
\affiliation{Centre for Theoretical Physics\\Jamia Millia Islamia, New Delhi 110025, India}

\begin{abstract}
We present a python package "ScalPy" for studying the late time scalar field cosmology for a wide variety of scalar field models, namely the quintessence, tachyon and Galileon model. The package solves the autonomous system of equations for power law and exponential potential. But it can be easily generalized to add more complicated potential. For completeness, we also include the standard parameterization for dark energy models, e.g. the $\Lambda$CDM, $w$CDM, $w_{0}w_{a}$CDM as well as the GCG parameterization.
The package also solves the linear growth equation for matter perturbations on sub-horizon scales. All the important observables related to background universe as well as to the perturbed universe, e.g. luminosity distance ($D_{L}(z)$), angular diameter distance ($D_{A}(z)$), normalized Hubble parameter ($h(z)$), lookback time ($t_{L}$), equation of state for the dark energy ($w(z)$), growth rate ($f=\frac{d \ln\delta}{d \ln a}$), linear matter power spectra ($P(k)$), and its normalization $\sigma_{8}$ can be obtained from this package. The code is further integrated with the publicly available MCMC hammer ``emcee" to constrain the different models using the presently available observational data. The code is available online at \url{https://github.com/sum33it/scalpy}
\end{abstract}
\maketitle
\section{Introduction}

There is compelling evidence that the universe is currently undergoing a phase of accelerated expansion \cite{scp, riess, planck, wmap9, sdss, boss, wigglez}. This indicates some new physics at cosmological scales (``dark energy") \cite{de} that can give rise to repulsive gravity resulting the universe to accelerate. The cosmological constant with $w = \frac{p}{\rho} = -1$ is the simplest example of repulsive gravity and is consistent with the majority of cosmological observations. But the observational data at present is also consistent  with dark energy behavior which is strictly not constant but evolves slowly throughout the cosmological history. Moreover the cosmological constant comes with its own baggage of problems like cosmic coincidence problem which can be avoided using dynamical dark energy. 

Scalar fields are the most widely used dynamical dark energy models where late time acceleration can be obtained by adjusting the slope of the scalar field potential around suitable epoch \cite{quint}. But the cosmological evolution of these models is severely constrained by very accurate cosmological observations. From the measurements of temperature anisotropy in the cosmic microwave background (CMB) radiation, the distance to last scattering is very well determined. This restricts the equation of state for the scalar field to be very close to $w=-1$ at present. The type-Ia supernovae observations as well as Baryon Acoustic Oscillation (BAO) measurement in the large scale structures also give similar constraint on the scalar field equation of state. On the other hand CMB measurements also restricts the scalar field contribution to the total energy density of the universe during recombination to be less than $1\%$. All these constraints disfavor \cite{linder}  the original ``tracker" kind of quintessence model\cite{lindcald}  where the scalar field initially has a fast roll phase before settling to the cosmological constant behavior at present. Just recently, analysis of the spectrum from a distant quasar finds no evidence of deviation in molecular lines produced around 12 billion years ago. This confirms no change in the mass ratio of the proton to the electron. The results also indicate that a dark energy scalar field—if it exists—has not evolved appreciably over $90\%$ the age of the Universe \cite{Bagdon}, thereby putting doubt on the tracker kind of models as a possible candidate for dark energy.

The other scalar field behavior that is consistent with the present observational data is the ``thawing" one \cite{lindcald}. Currently there are many types of scalar field models that can give rise to thawing behavior. Examples are quintessence \cite{quint}, tachyons \cite{tach} as well as Galileon \cite{gal} models. In all these thawing models, the thawing nature does not depend crucially on the form of the potential. Once the initial conditions are chosen in such way that the scalar field starts in a nearly frozen state with $w\approx -1$, the thawing behavior is ensured. Fine tuning this initial condition is an issue in all thawing type of models. In a recent investigation, attempt has been made to relate this initial condition to inflationary dynamics \cite{raghu}. 

Given the fact that thawing models are observationally promising as a possible candidate for dynamical dark energy models, we need to have a combined setup to investigate the cosmological evolution for different thawing scalar field models. In this work, we present such a setup in a Python environment. We name this setup as ``ScalPy" which solves the autonomous system with a given scalar field model by providing suitable initial conditions and calculates the observables like normalized Hubble parameter, angular diameter distance, luminosity distance, growth function, growth rate, linear matter power spectrum etc. We consider two simple potentials , the power-law and exponential for all the different versions of the scalar fields as these are the two most widely used potentials for dark energy in the literature. But one can easily incorporate the other potentials in this package. We also combine the package with the publicly available MCMC sampler ``emcee" \cite{ForemanMackey:2012ig} to do the statistical analysis of different scalar fields with observational data. 

Given the promise of thawing scalar field models as possible candidate for the dark energy, there are several investigations related to the thawing model for different types of scalar field. All these works are related to individual scalar field models and there are not many attempts to study the different thawing scalar field model in a combined approach. This is the first such combined approach in python environment.

This paper is structured as follows. In the section II, we describe the autonomous system of equations for different scalar field models and also the method to fix the initial conditions to solve these systems; in section III, we describe different cosmological observable that ScalPy calculates for different scalar field models; in section IV, we describe different data sets that are used to put observational constraints on different models and the results of the data analysis; finally we summarizes our results in the conclusion section. In the Appendix part of the manuscript, we briefly describe how to install and to use ScalPy. 

\section{Dynamics of cosmological scalar fields}

In the present investigation, we confine ourselves to flat FRW universe with scale factor $a(t)$ and given by the line element

\begin{equation}
 ds^2 = -dt^2 + a^2(t)(dr^2 + r^2d\Omega^2),
\end{equation}

\noindent
where $d\Omega^2 = (d\theta^2 + sin^2(\theta)d\phi^2)$. The scale factor at present is set to be $a_{0} =1$.  In the late time evolution of the universe, the dominant contribution to the energy budget of the universe is from the non-relativistic matter (baryons+dark matter) and from the dark energy. In the present paper we are interested in studying various forms of scalar field models as the dark energy component in the universe..

In following subsection we will look at quintessence, tachyon and Galileon fields and see how we make autonomous systems in these frameworks.

\subsection{Quintessence}
We consider a minimally coupled scalar field and Lagrangian for such a field is given by
\begin{equation}
 \mathcal{L} = \frac{1}{2}\partial_\mu\phi\partial^\mu\phi - V(\phi),
\end{equation}

\noindent
where $V(\phi)$ is the given potential. One can calculate the energy momentum tensor and by considering perfect fluid, the energy density and pressure of quintessence field is given by
\begin{eqnarray}
 \rho_\phi &=& \frac{1}{2}\dot{\phi}^2 + V(\phi),\\
 P_\phi &=& \frac{1}{2}\dot{\phi}^2 - V(\phi).
\end{eqnarray}
\noindent
The equation of motion for scalar field obtained by variation of action with respect to field $\phi$ is
\begin{equation}
 \ddot{\phi} + 3H\dot{\phi} + V_\phi = 0.
\end{equation}

\noindent
where the subscript $\phi$ denotes the derivative with respect to field $\phi$. The Hubble equation is given by

\begin{equation}
 H^2 =  \frac{1}{3 M^2_{\rm{pl}}}\rho,
\end{equation}

\noindent
where $\rho = \rho_m + \rho_\phi$ is the total density of matter and scalar field. We define the variables $x$, $y$, and $\lambda$ as \cite{quint}

\begin{align}
x&=\frac{\phi^\prime}{\sqrt{6} M_{\rm{pl}}}\,,\quad y=\frac{\sqrt{V}}{\sqrt{3} H M_{\rm{pl}}},\\
\lambda &= -M_{\rm{pl}} \frac{V_\phi}{V}\,, \quad \Gamma = V\frac{V_{\phi\phi}}{V_{\phi}^2}.
\end{align}

\noindent
Here the prime denote the derivative with respect to the e-folding $N$ (= $\ln a$) i.e. $\phi^\prime \equiv a(d\phi/da)$ and subscript `$\phi$' denote the derivative with respect to field $\phi$. In terms of these variables

\begin{equation}
\Omega_\phi = x^2 + y^2,\\
\gamma \equiv 1+w = \frac{2x^2}{x^2 + y^2},
\end{equation}
\noindent
where $\Omega_\phi$ is the density parameter for scalar field and $w = \frac{P}{\rho}$ is the equation of state. We can construct an autonomous system using variables $x$, $y$ and $\lambda$

\begin{eqnarray}
x^\prime &=& -3x + \lambda\sqrt{\frac{3}{2}}y^2 + \frac{3}{2}x[1 + x^2-y^2]\label{quint1}\\
y^\prime &=& -\lambda\sqrt{\frac{3}{2}}xy + \frac{3}{2}y[1+x^2-y^2] \label{quint2}\\
\lambda^\prime &=& - \sqrt{6} \lambda^2(\Gamma - 1) x. \label{quint3}
\end{eqnarray}

In terms of $\Omega_{\phi}$, $\gamma$ and $\lambda$, the above equations becomes \cite{ss},
\begin{eqnarray}
\gamma^\prime &=& -3\gamma(2-\gamma) + \lambda(2-\gamma)\sqrt{3 \gamma
\Omega_\phi},\label{quint4}\\
\Omega_\phi^\prime &=& 3(1-\gamma)\Omega_\phi(1-\Omega_\phi),\label{quint5}\\
\lambda^\prime &=& - \sqrt{3}\lambda^2(\Gamma-1)\sqrt{\gamma \Omega_\phi}. \label{quint6}
\end{eqnarray}

\subsection{Tachyons}

Cosmological dynamics for tachyonic field can be obtained by the Dirac-Born-Infeld (DBI) type action \cite{tach}
\begin{equation}
 S_{tachyon} = -\int d^4x V(\phi) \sqrt{-g} \sqrt{1 - \partial^\mu\phi\partial_\mu\phi}.
\end{equation}
In natural units, the dimension of the tachyon field is $[Mass]^{-1}$. The energy density, pressure and equation of state of the tachyon field is given by
\begin{eqnarray}
 \rho_\phi &=& \frac{V(\phi)}{\sqrt{1 - \dot{\phi}^2}},\\
 P_\phi &=& -V(\phi) \sqrt{1 - \dot{\phi}^2},\\
 w &=& - (1 - \dot{\phi}^2).
\end{eqnarray}

\noindent
The equation of motion for Tachyon field is
\begin{equation}
 \ddot{\phi} + 3H\dot{\phi}(1-\dot{\phi}^2) + \frac{V'}{V}(1-\dot{\phi}^2).
\end{equation}

\noindent
Again we define following dimensionless variables
\begin{align}
 x &=H\phi' \,,\quad y =\frac{\sqrt{V}}{\sqrt{3}H M_{\rm{pl}}},\\
 \lambda &=-M_{\rm{pl}}\frac{V_\phi}{V^{3/2}} \,,\quad \Gamma = V \frac{V_{\phi\phi}}{V_{\phi}^2}.
\end{align}

\noindent
In terms of these variables, equation of state $w_\phi$ and density parameter $\Omega_\phi$ for the tachyon field is
\begin{equation}
 \gamma = (1+w_\phi) = x^2 \, , \quad \Omega_\phi = \frac{y^2}{\sqrt{1 - x^2}}.
\end{equation}

\noindent
We construct the autonomous system for tachyon field \cite{Ali:2009mr}
\begin{align}
 \gamma' &= -6\gamma(1-\gamma) + 2\sqrt{3\gamma\Omega_\phi}\lambda(1 - \gamma)^{5/4}, \label{tachyon1}\\
 \Omega_\phi' &= 3\Omega_\phi(1-\gamma)(1 - \Omega_\phi),\label{tachyon2}\\
 \lambda' &= -\sqrt{3\gamma\Omega_\phi}\lambda^2(1 - \gamma)^{1/4}(\Gamma - 3/2),\label{tachyon3}.
\end{align}

\subsection{Galileon}
The large scale modification of gravity which involves an effective scalar field $\pi$ can explain late time acceleration of the Universe. This field $\pi$ is called ``Galileon" and its Lagrangian respects the shift symmetry in the Minkowski background satisfying $\pi \rightarrow \pi + c$ and $\partial_\mu\phi \rightarrow \partial_\mu\pi + b_\mu$ where $c$ and $b_\mu$ are constants \cite{gal}. The action for Galileon field up to third order is \cite{wali}
\begin{equation}
S=\int d^4x\sqrt{-g}\Bigl [\frac{M^2_{\rm{pl}}}{2} R -  \frac{1}{2}(\nabla \pi)^2\Bigl(1+\frac{\alpha}{M^3}\Box \pi\Bigr) - V(\pi) \Bigr] + \mathcal{S}_m,
\label{1.1}
\end{equation}
where $M_{\rm{pl}} = \sqrt{\frac{1}{8\pi G}}$ is reduced Planck Mass. $\alpha$ is the dimensionless constant and setting it equal to zero givers the standard quintessence action. $M$ is the constant with mass dimension and it can be fixed to $M_{\rm{pl}}$ by redefining $\alpha$. Again by defining following dimensionless variables,
\begin{align}
x&=\frac{\dot{\pi}}{\sqrt{6}H M_{\rm{pl}}}\,,\quad y=\frac{\sqrt{V}}{\sqrt{3} H M_{\rm{pl}}},\\
\epsilon &=-6\frac{\alpha}{M_{\rm{pl}}^3}H\dot \pi\,, \quad \lambda=-M_{\rm{pl}}\frac{V_\pi}{V} \, , \quad \Gamma= V \frac{V_{\pi\pi}}{V_{\pi}^2},
\end{align}
we construct the autonomous system \cite{wali}
\begin{widetext}
\begin{align}
x'&=\frac{3 x^3 \left(2+5 \epsilon +\epsilon ^2\right)-3 x \left(2-\epsilon +y^2 (2+3 \epsilon )\right)+2 \sqrt{6} y^2 \lambda -\sqrt{6} x^2 y^2 \epsilon  \lambda }{4+4 \epsilon +x^2 \epsilon ^2}, \label{galileon1}\\
y'&=-\frac{y \left(12 \left(-1+y^2\right) (1+\epsilon )-6 x^2 \left(2+4 \epsilon +\epsilon ^2\right)+\sqrt{6} x^3 \epsilon ^2 \lambda +2 \sqrt{6} x \left(2+\left(2+y^2\right) \epsilon \right) \lambda \right)}{8+8 \epsilon +2 x^2 \epsilon ^2},\label{galileon2}\\
\epsilon' &=-\frac{\epsilon  \left(-3 x \left(-3+y^2\right) (2+\epsilon )+3 x^3 \left(2+3 \epsilon +\epsilon ^2\right)-2 \sqrt{6} y^2 \lambda -\sqrt{6} x^2 y^2 \epsilon  \lambda \right)}{x \left(4+4 \epsilon +x^2 \epsilon ^2\right)},\label{galileon3}\\
\lambda' &=\sqrt{6}x\lambda^2(1-\Gamma),\label{galileon4}.
\end{align}
\end{widetext}
\noindent
For $\epsilon=0$ we recover the autonomous system for the standard quintessence \eqref{quint1}-\eqref{quint3}.

\noindent
The density parameter $\Omega_m$ and equation of state for the $\pi$ field is given by:
\begin{align}
\Omega_m &=1-x^2 (1+\epsilon)-y^2, \label{galileon5}\\
\omega_{\pi}&=\frac{-12 y^2 (1+\epsilon )+3 x^2 \left(4+8 \epsilon +\epsilon ^2\right)-2 \sqrt{6} x y^2 \epsilon  \lambda }{3 \left(4+4 \epsilon +x^2 \epsilon ^2\right) \left(y^2+x^2 (1+\epsilon )\right)}. \label{galileon6}
\end{align}

\subsection{Solving the dynamical systems}
To solve a cosmological dynamical system, we have to choose a point in time where we set the appropriate initial conditions and then let the system evolve till present day. We give initial conditions at around decoupling ($a \approx 10^{-3}$). 

In case of quintessence and tachyons, we solve the autonomous system of equations in terms of $\gamma$, $\Omega_{\phi}$ and $\lambda$ given by equations \eqref{quint4}-\eqref{quint6} and \eqref{tachyon1}-\eqref{tachyon3}for quintessence and tachyon respectively. We have to provide initial conditions for $\gamma_{i}$, $\Omega_{\phi i}$ and $\lambda_{i}$ where ``i'' represents the initial value. As we are considering the thawing class of models, the scalar field is initially nearly frozen due to large Hubble friction and hence we choose $\gamma_{i}$ to be close to zero. Throughout rest of the paper, we fix $\gamma_{i} = 10^{-4}$. $\lambda_{i}$ is the initial slope of the potential and $\lambda_{i} = 0$ gives an exact cosmological constant behavior. So values for $\lambda_{i}$ different from zero, gives the deviation from cosmological constant. In our subsequent calculations, we choose $\lambda_{i}$ to be a parameter of the model. As the contribution of dark energy to be negligible in the early time, $\Omega_{\phi i}$ should be close to zero initially. But it has to be fixed initially in such a way that for a given $\gamma_{i}$ and $\lambda_{i}$, the system evolves to present day with the appropriate value of $\Omega_{\phi 0}$ where ``0'' stands for present day value. We check that $\Omega_{\phi i}$ around $10^{-9}$ gives suitable values for $\Omega_{\phi 0}$. So in our calculations, we choose $\Omega_{\phi i}$ to be the second parameter and varies it between $1 \times 10^{-9}$ to $10 \times 10^{-9}$. 

For Galileon field, we solve autonomous systems in terms of variables $x$, $y$, $\epsilon$ and $\lambda$ given by the equations \eqref{galileon1}-\eqref{galileon4} and hence we have to provide four initial conditions for  $x_i$, $y_i$, $\epsilon_i$ and $\lambda_i$ respectively. As here also, we consider the thawing behavior only, the Galileon field is initially nearly frozen and hence $x_{i} \approx 0$. We set $x_i = 10^{-7}$ in our calculations. For $\Omega_{\phi i}$ to be the same as in quintessence and tachyon, $\epsilon_{i}$ and $y_{i}$ are now related by equation \eqref{galileon5}. So we vary $\epsilon_{i}$ and $\Omega_{\phi i}$ as two parameters and $y_{i}$ is fixed subsequently by equation \eqref{galileon5}. We also consider $\lambda_{i}$ as our third parameter as it gives the deviation from cosmological constant. So in Galileon models, we have three parameters, $\Omega_{\phi i}, \epsilon_{i}$ and $\lambda_{i}$.

For power law and exponential potential, $\Gamma$ is a constant while for other potentials, $\Gamma$ evolve as a function of $\lambda$ in equations \eqref{quint3},\eqref{tachyon3} and \eqref{galileon4}.
\begin{figure*}[t]
    \centering
    \subfigure
    {
        \includegraphics[scale=0.4]{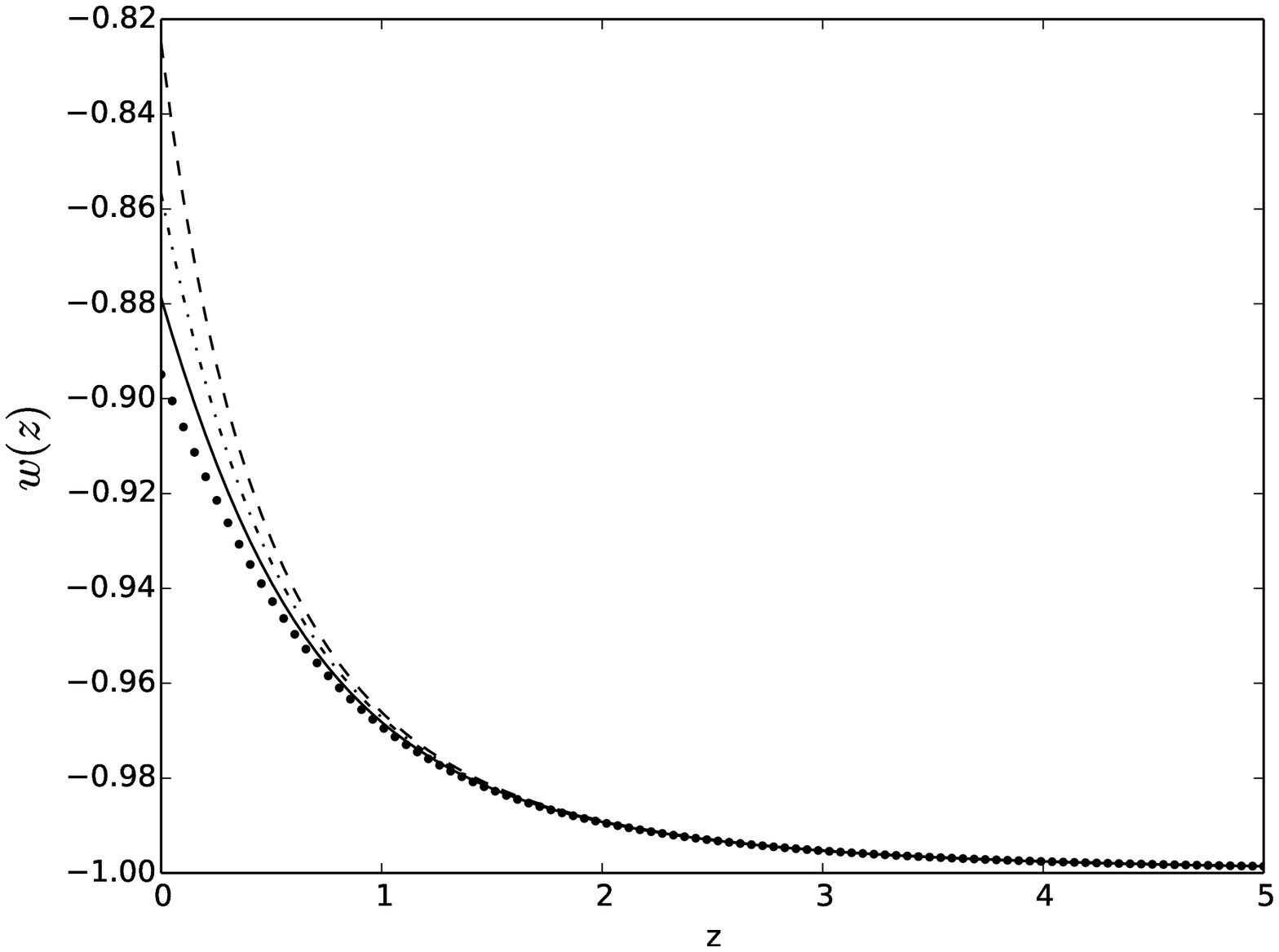}
    }
    \\
    \subfigure
    {
        \includegraphics[scale=0.4]{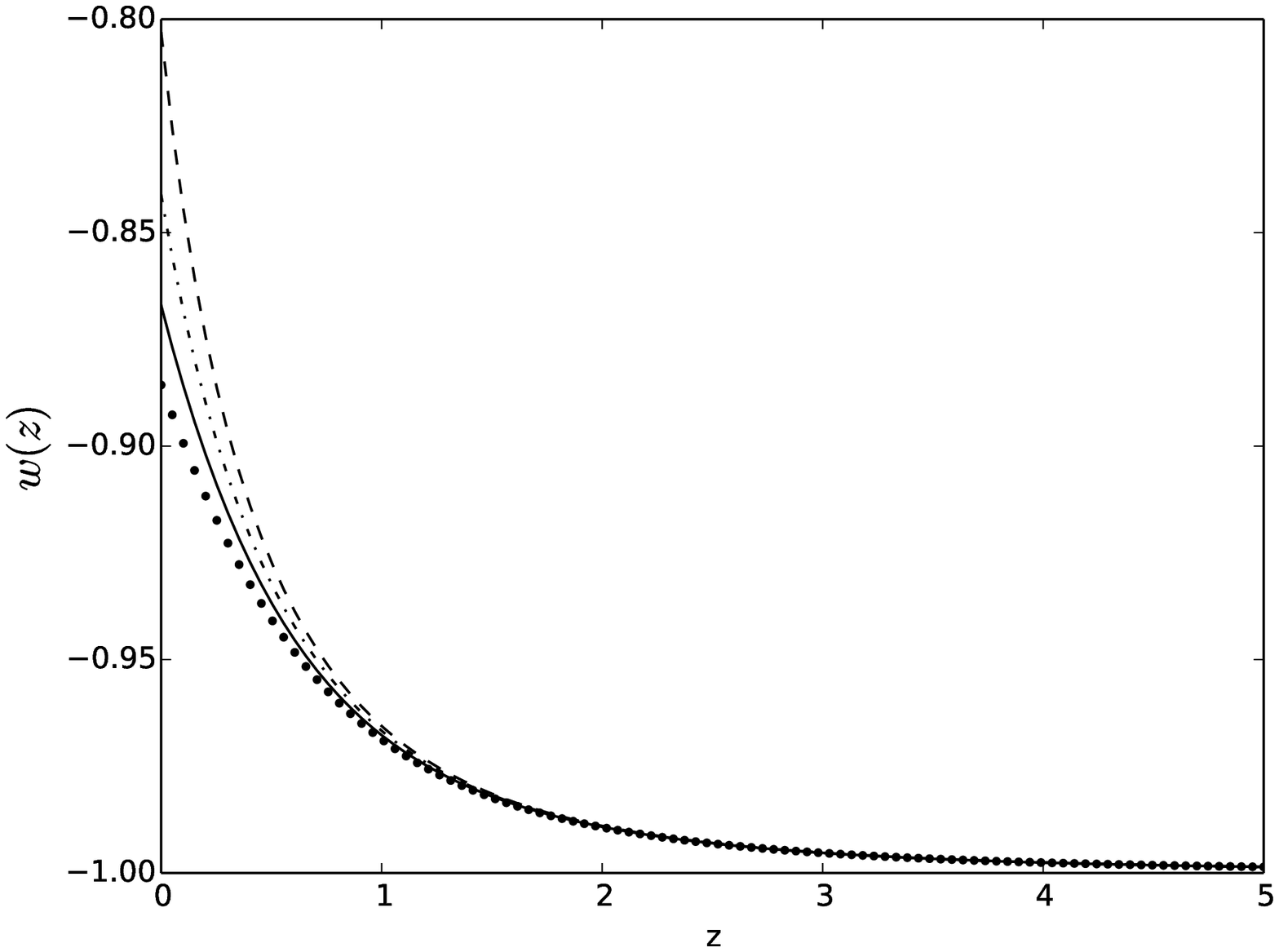}
    }
    \subfigure
    {
        \includegraphics[scale=0.4]{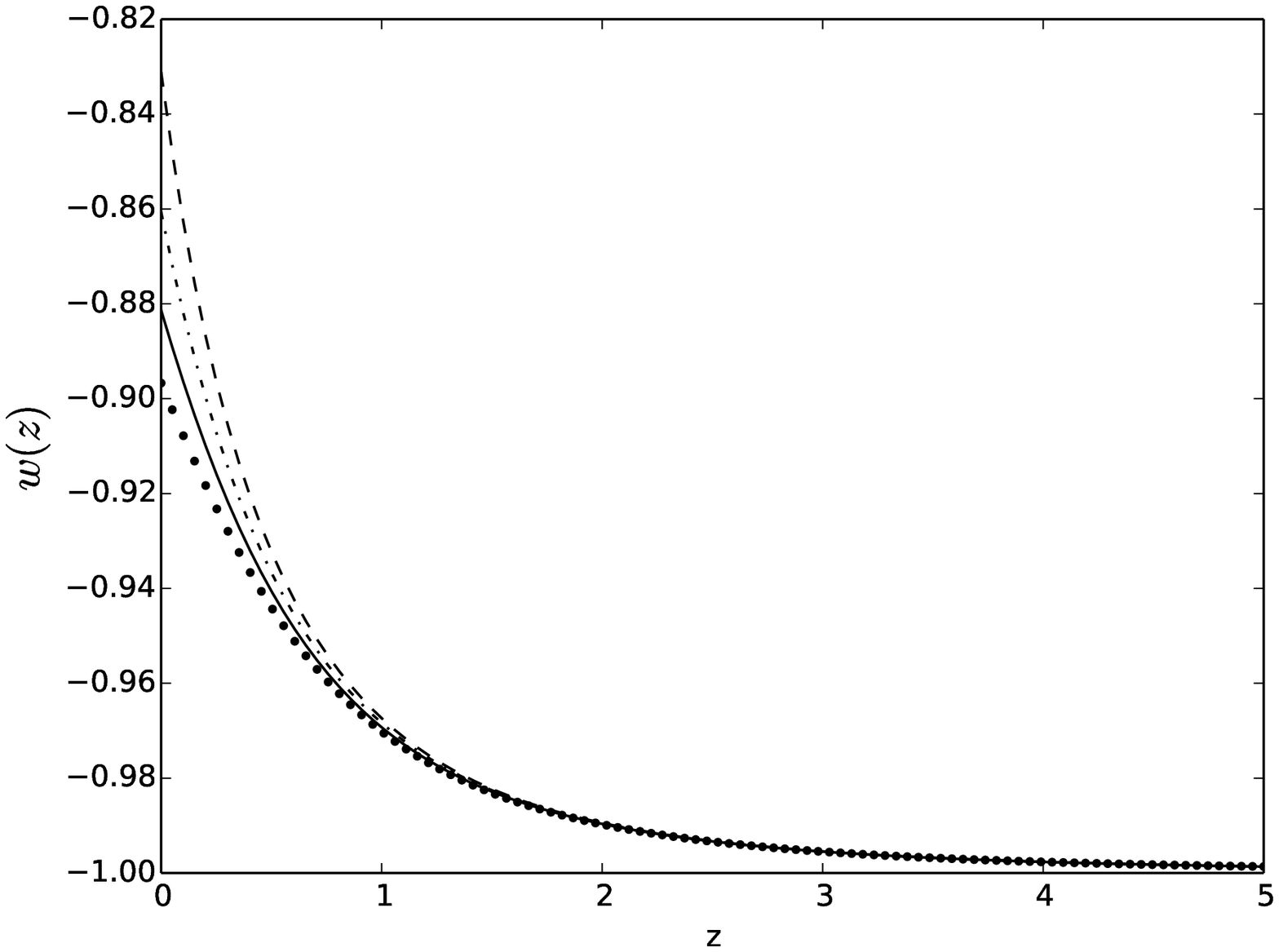}
    }
    \caption{Equation of state $w$ as a function of redshift z for quintessence field (top), tachyon field (bottom left) and Galileon field (bottom right). Solid lines corresponds to exponential potential while dashed, dotted-dashed and dotted lines corresponds to power law potential with $n = 1,2$ and $-1$ respectively.}
\end{figure*}

\section{cosmological Observables}
Following David Hogg \cite{Hogg:1999ad} we define observable quantities such as Hubble distance, luminosity distance, angular diameter distance and lookback time. The Hubble distance is defined as $D_H = \frac{c}{H_0}$ where c is the speed of light and $H_0$ is the Hubble constant. Since we have assumed a flat FLRW metric, the line of sight comoving distance is defined as
\begin{equation}
 D_c = \frac{c}{H_0}\int_{0}^{z}\frac{dz'}{h(z')},
\end{equation}
\noindent
where $h(z)=H(z)/H_0$ is normalized Hubble parameter. The angular diameter distance $D_A$ and luminosity distance $D_L$ are related to comoving distance as
\begin{equation}
 D_L = (1+z)D_m = (1+z)^2D_A,
\end{equation}
\noindent
where $D_m$ is transverse comoving distance which is equal to $D_c$ for the flat Universe. 
\noindent
We also define lookback time $t_L$ to an object as the difference between age of the Universe and the time when the light was emitted from that object:
\begin{equation}
 t_L = \frac{1}{H_0}\int_0^{z}\frac{dz'}{(1+z')h(z')}.
\end{equation}
\noindent
The age of the Universe is given by,
\begin{equation}
 t = \frac{1}{H_0}\int_0^{\infty} \frac{dz'}{(1+z')h(z')}.
\end{equation}

\noindent
We also define Om diagnostic \cite{Sahni:2008xx} in our package :
\begin{equation}
 Om(x) = \frac{h^2(x) - 1}{x^3 -1} ;\quad x = 1+z.
\end{equation}

\noindent
The time delay distance used in gravitational lensing cosmography is also included in the package. It is defined as \cite{Suyu:2012aa}
\begin{equation}
 D_{\vartriangle t} = (1 + z_d)\frac{D_d D_s}{D_{ds}}.
\end{equation}
Where $D_d$ is the angular diameter distance to the deflector or lens at redshift $z_d$ and $D_s$ is the angular diameter distance to the source. $D_{ds}$ is the angular diameter distance between source and deflector.

In figure 1 we plotted equation of state, $w$ as a function of redshift for each type of scalar field (quintessence, tachyon and Galileon) with power law potential ($n = 1,2,-1$) and for exponential potential. The initial conditions we choose for quintessence and tachyon fields are ($\Omega_{\phi i},\lambda_i$) $\sim$ ($2.5\times10^{-9}, 0.9$) and for Galileon field we choose ($\Omega_{\phi i}, \epsilon_i,\lambda_i$) $\sim$ ($2.5\times10^{-9}, 2.0, 0.9$). The value of $\gamma_i$ for all three types of field is taken to be $10^{-4}$.

\begin{figure*}[t]
    \centering
    \subfigure
    {
        \includegraphics[scale=0.4]{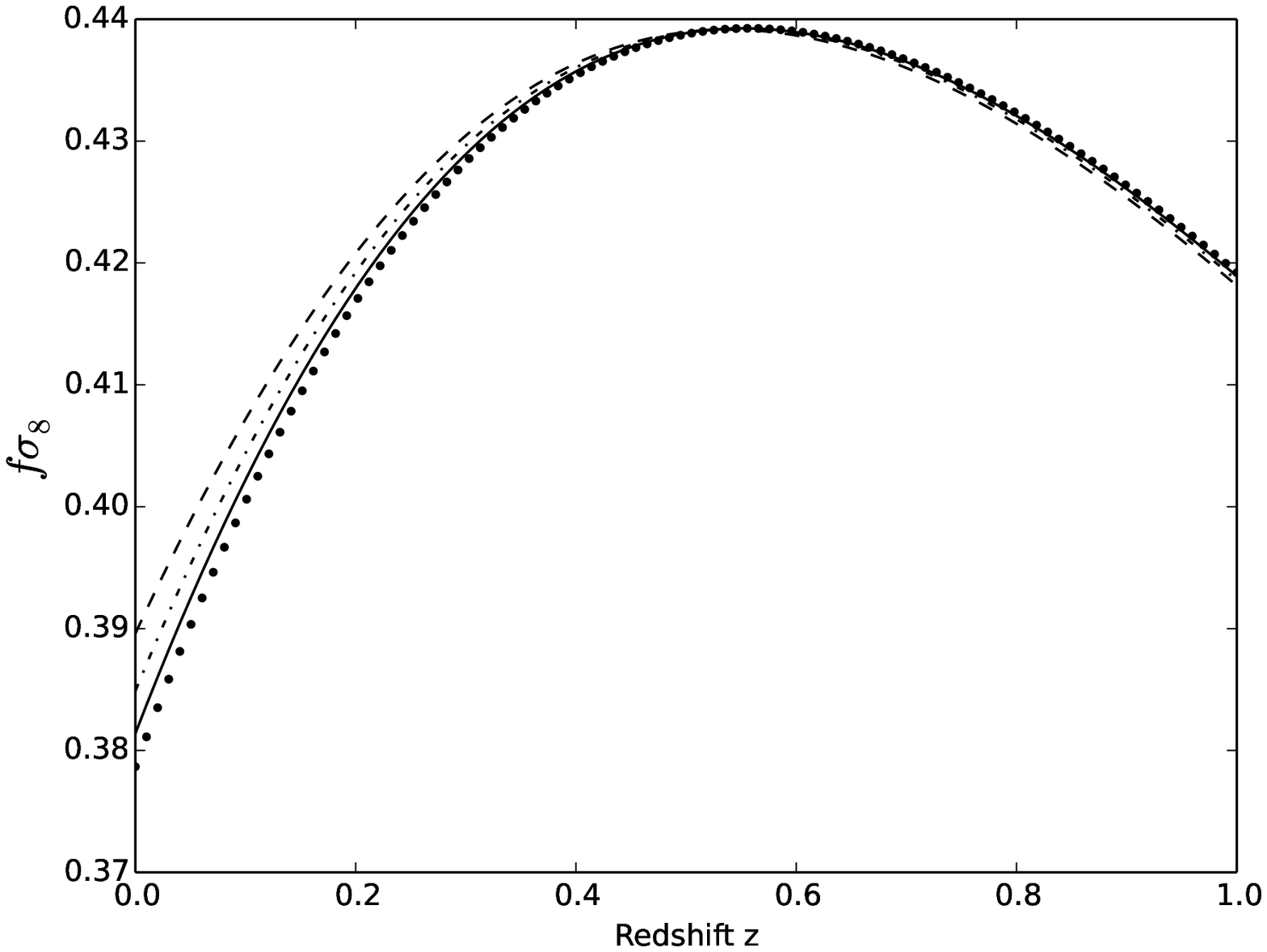}
    }
    \\
    \subfigure
    {
        \includegraphics[scale=0.4]{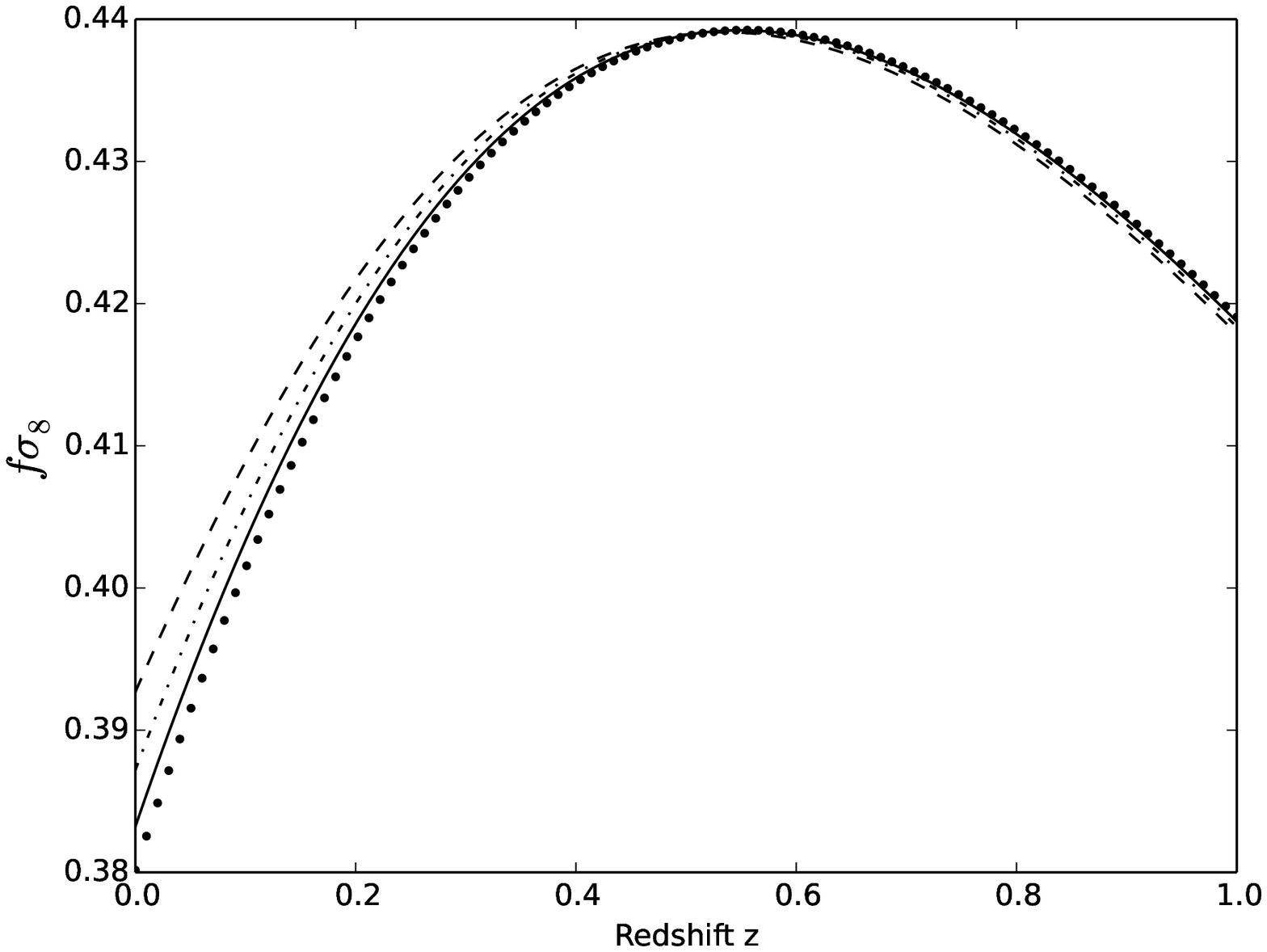}
    }
    \subfigure
    {
        \includegraphics[scale=0.4]{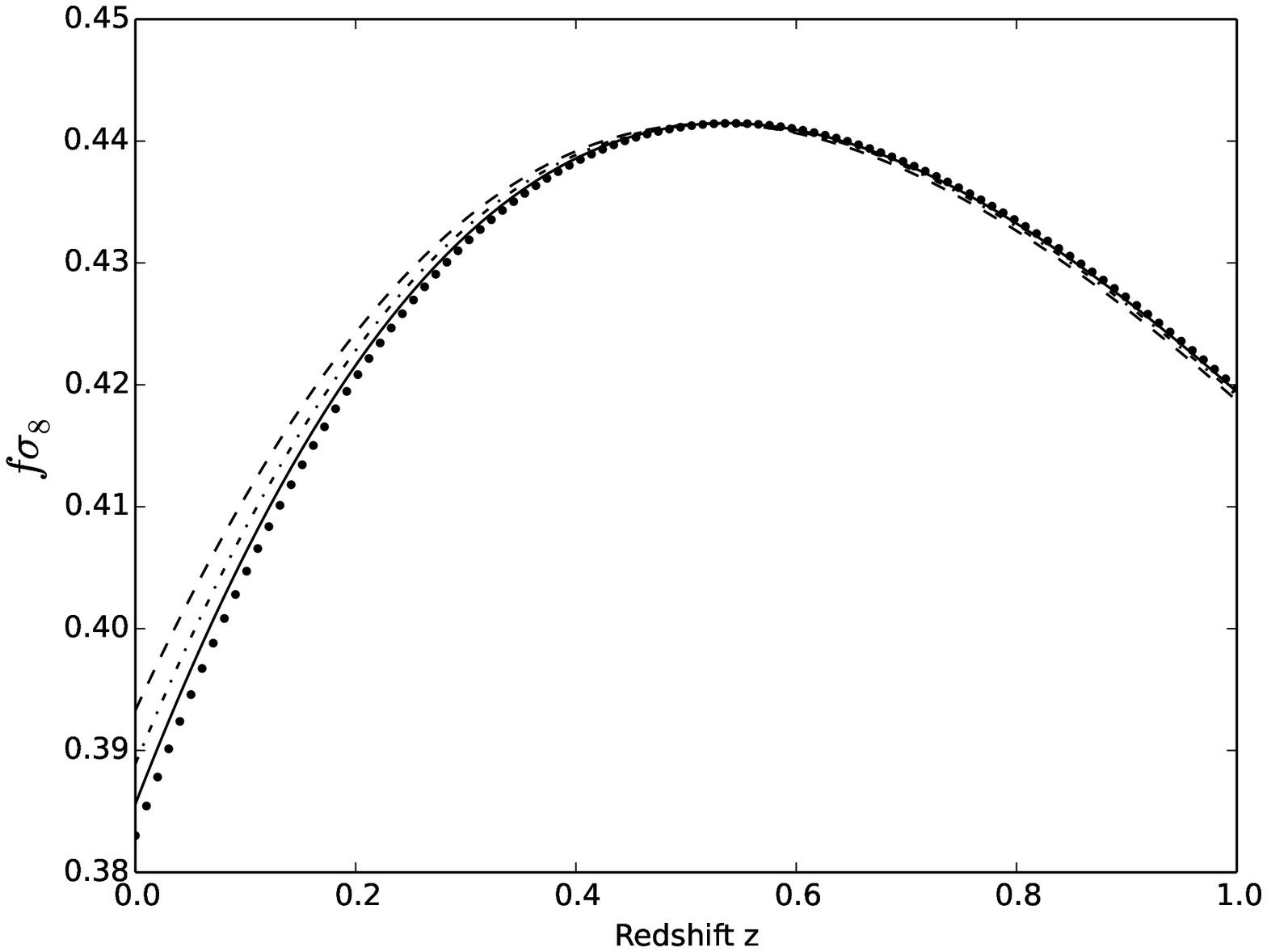}
    }
    \caption{$f\sigma_8$ redshift z for scalar field (top), tachyon field (bottom left) and Galileon field (bottom right). Solid lines corresponds to exponential potential while dashed, dotted-dashed and dotted lines corresponds to power law potential with $n = 1,2$ and $-1$ respectively..}
\end{figure*}

\subsection{Growth of inhomogeneities}
We consider perturbation in matter part only and scalar field is homogeneous. We are interested in structure formation to scale much smaller than the Hubble radius so the Newtonian approximation is valid. The density contrast of matter field $\delta_m = \frac{\overline{\rho}_m - \rho_m}{\overline{\rho}_m}$ follows the equation at sub-horizon approximation

\begin{equation}
 \ddot{\delta}_m + 3H\dot{\delta}_m + 4\pi G\rho_m \delta_m = 0.
\end{equation}

Taking Fourier transform of above equation and defining the linear growth function $D$ and the linear growth rate $f$ as

\begin{eqnarray}
\delta_{k}^{lin} (a) &\equiv& D(a)\delta_{k}^{ini},\\
f &=& \frac{d \ln{D}}{d \ln{a}}.
\end{eqnarray}

We calculate the linear matter power spectrum defined as

\begin{equation}
P_{lin}(z,k) = A_0 k^{n_s} T^2(k) D^2_{n}(z).
\end{equation}

\noindent
Here $A_0$ is the normalization constant, $n_{s}$ is spectral index for the primordial density fluctuations generated through inflation, $D_n(z)$ is growth function normalized such as it is equal to unity at $z=0$ i.e. $D_n(z)=\frac{D(z)}{D(0)}$ and $T(k)$ is the transfer function as prescribed by Eisenstein and Hu \cite{eisenhu}
for a mixture of CDM and baryons:

\begin{equation}
T(k) = \left(\frac{\Omega_{b0}}{\Omega_{m0}}\right)T_{b}(k) + \left(\frac{\Omega_{c0}}{\Omega_{m0}}\right)T_{c}(k),
\end{equation}

\noindent
where $\Omega_{c0}$ is density parameter for CDM and $\Omega_{m0} = \Omega_{c0}+\Omega_{b0}$. 
The form for $T_{b}(k)$ and $T_{c}(k)$ are given by Eisenstein and Hu \cite{eisenhu}. An important quantity $\sigma_8$ is defined as the fluctuation of mass within the boxes of $8h^{-1}$ Mpc when we move from place to place in present day universe.:

\begin{equation}
\sigma^{2}(a,R) = \int^{\infty}_{0} \Delta^{2}(k,a) W^{2}(k,R) d \ln{k},
\end{equation}

\noindent 
where the window function $W(k,R)$  is defined as $W(k,R) = 3\left( \frac{\sin(KR)}{(kR)^3} + \frac{\cos(kR)}{(kr)^2}\right)$ and for $\sigma_{8}$, one puts $ R = 8h^{-1}$ Mpc.

Using definition of $\sigma_8$ and window function, we fix the normalization constant $A_0$ as \cite{Knobel:2012wa}
\begin{equation}
 A_0 = 2\pi^2\sigma_8^2\left(\int k^{n_s + 2} T^2(k) W^2(k,R) dk \right)^{-1},
\end{equation}
\noindent
where we take $R = 8h^{-1}$ Mpc. Fig 2 shows the behavior of $f\sigma_8$ as a function of z for quintessence, tachyons and Galileon fields with exponential as well as power law potential. We used the same initial conditions as used for plotting equation of state in figure 1.

\begin{figure*}[t]
    \centering
    \subfigure
    {
        \includegraphics[scale=0.4]{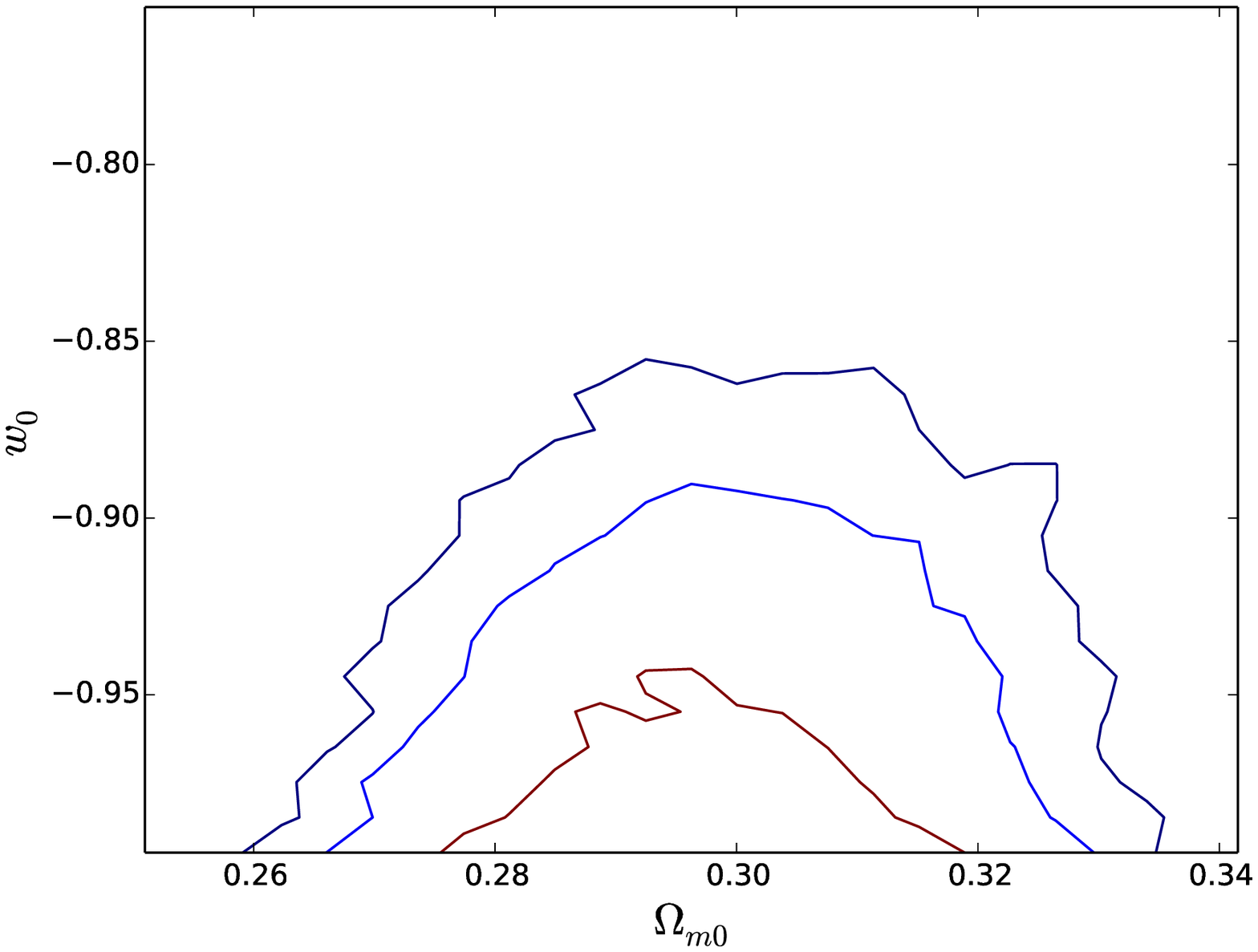}
    }
    \\
    \subfigure
    {
        \includegraphics[scale=0.4]{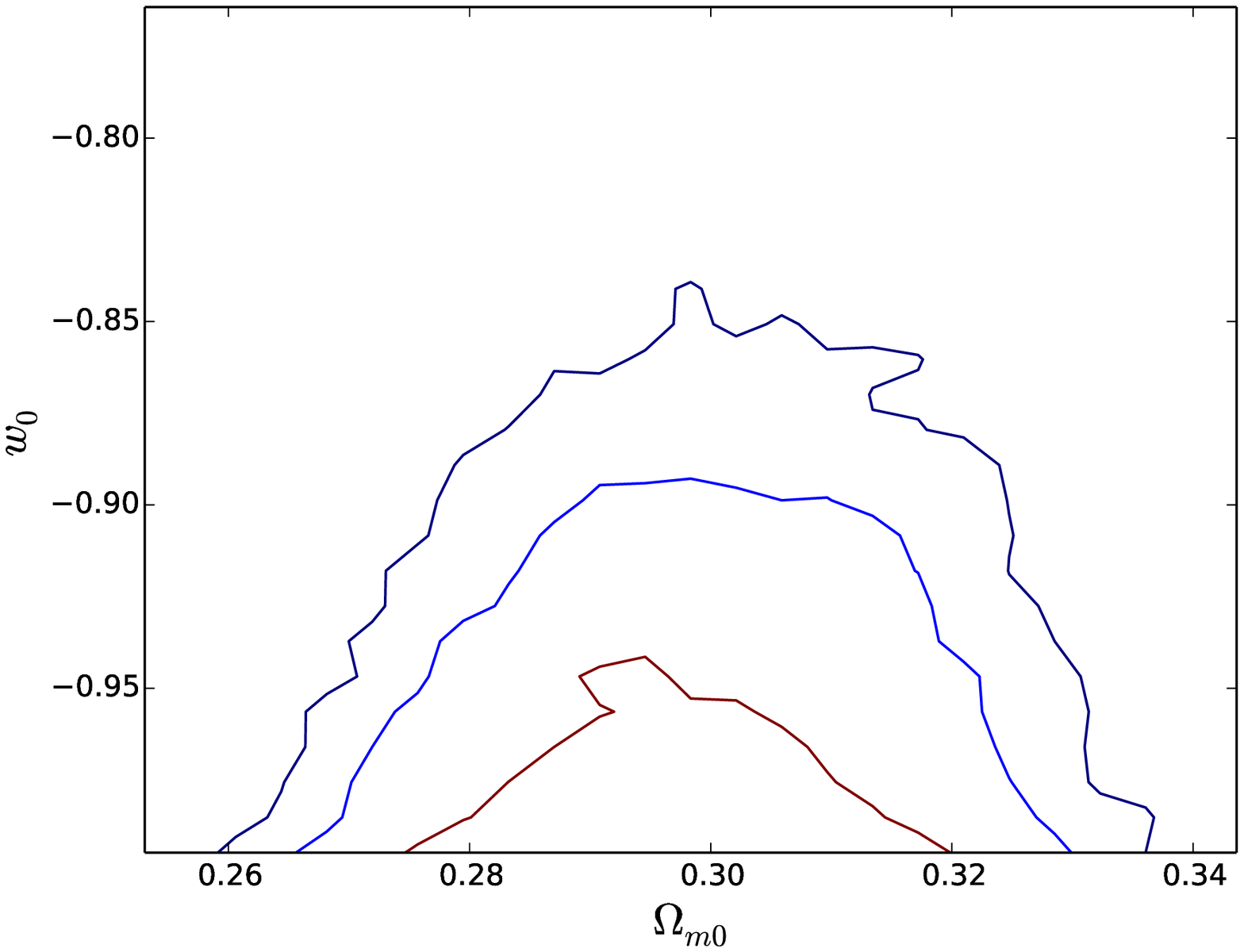}
    }
    \subfigure
    {
        \includegraphics[scale=0.4]{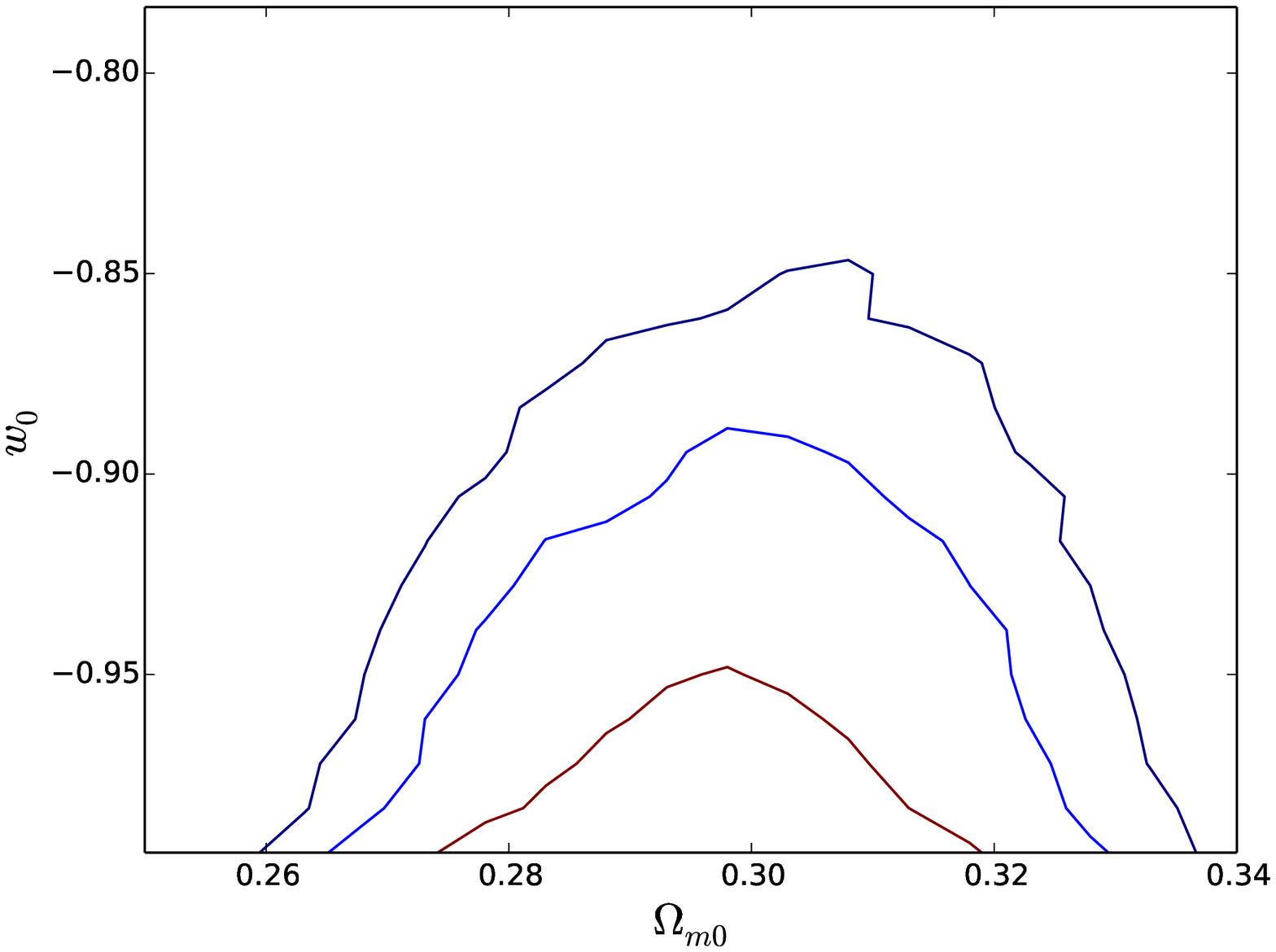}
    }
    \caption{1$\sigma$, 2$\sigma$ and 3$\sigma$ contours obtained for $JLA+BAO+Shift(from CMB)$ for quintessence field (top), tachyon (bottom left) and Galileon field (bottom right) with linear potential.}
\end{figure*}

\section{Data Analysis}
We used our ScalPy package and MCMC hammer `emcee' \cite{ForemanMackey:2012ig} to constrained model parameters of quintessence, tachyon and Galileon fields as well as $w$CDM, $w_0w_a$CDM and $GCG$ models(See Appendix). We used following data sets for our analysis:

\begin{itemize}
 \item 31 binned distance modules fitted to JLA sample given by by M. Betoule et al. \cite{Betoule:2014frx}
 \item  Combined BAO/CMB constraints on the angular scales of the BAO oscillations in the matter power spectrum measured by SDSS survey, 6dF Galaxy survey and the Wiggle-z survey. We used covariance matrix for this provided by Giostri et al \cite{Giostri:2012ek}.
 \item We also used measurement of shift parameter of CMB as given in Shafer et al \cite{Shafer:2013pxa}.
 \item Measurements for $f\sigma_{8}$ by various Galaxy surveys like 2dF, SDSS, 6dF, BOSS and Wiggle-Z. A compilation of different measurements for $f\sigma_{8}$ has been provided by Basilakos et al.\cite{Basilakos:2013nfa}
\end{itemize}
The results for data analysis for quintessence, tachyon field and Galileon field with linear potential are shown in figure (3). We used $h=0.7$ in our analysis. We show the 1$\sigma$, 2$\sigma$ and 3$\sigma$ contours in $\Omega_{\phi}-w$ plane. The $\Lambda$CDM behavior ($w = -1$) is allowed for all these types of scalar fields within 1$\sigma$ region with considerable range of $\Omega_{m0}$ values. One can also see that the constraints on present day equation of state $w_0$ and density parameter for matter $\Omega_{m0}$ are same for all three type of scalar fields with linear potential. This result also holds for other power law potentials.

In figure (4) and (5) we show the constraints on model parameters for $w_0w_a$CDM and $w$CDM models respectively. The concordance $\Lambda$CDM model is within $1\sigma$ error bar when we combine all the four data sets used. The constraints on the model parameters of GCG model is shown in figure (6). The parameter $A_s$ is related to present day equation of state $w_0$ as $w_0 = -A_s$. We see that $w_0 = -1$ is again allowed within $1\sigma$ region for GCG as well.

\begin{figure}
 \begin{center}
 \begin{tabular}{cc}
  & \\
 {\includegraphics[width=2.6in,height=2.4in,angle=0]{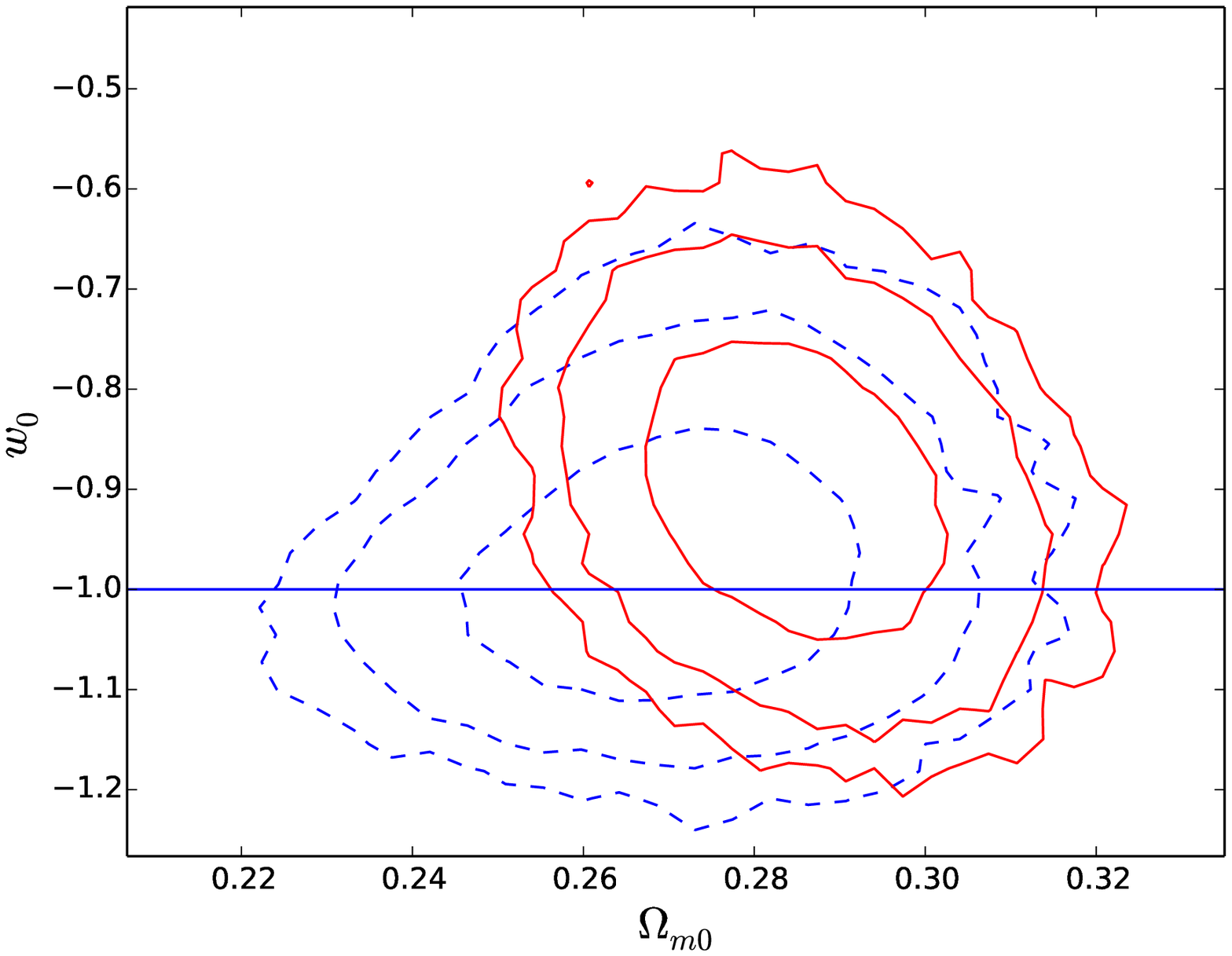}}&
 {\includegraphics[width=2.6in,height=2.4in,angle=0]{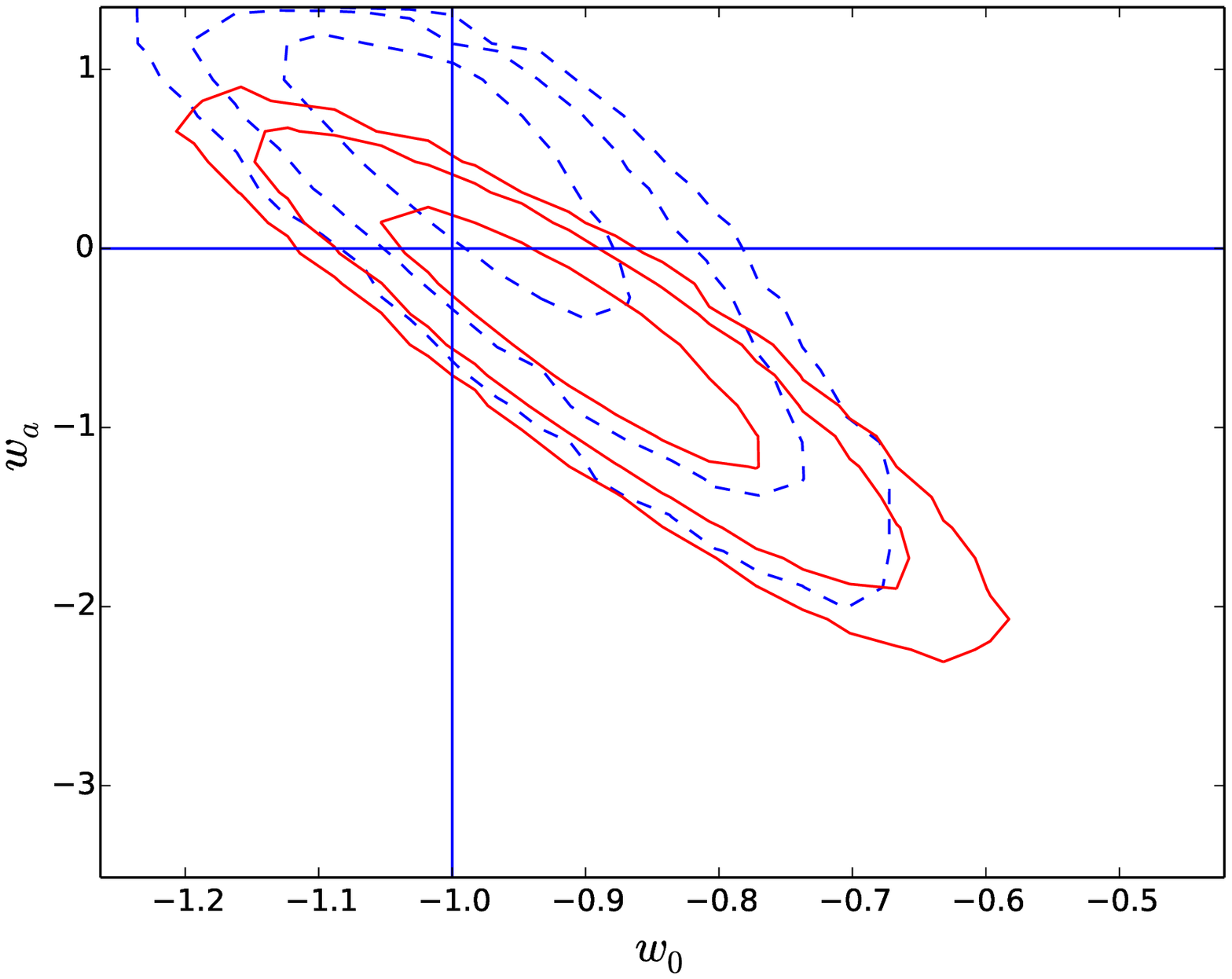}}
 \\

 \end{tabular}
 \caption{1$\sigma$, 2$\sigma$ and 3$\sigma$ contours obtained for $JLA+BAO+CMB~Shift$ (dashed) and $JLA+BAO+CMB~Shift+f\sigma_8$ (solid) on model parameters for $w_0w_aCDM$ model.}
 \end{center}
 \end{figure}

\section{Conclusion}
To summarize, we present a python package ScalPy to study the late time cosmology with different kind of scalar field models. To begin with, we consider only the minimally coupled scalar field and a flat FRW universe. We also neglect the radiation component as it will not affect the results in the late time. But one can trivially incorporate radiation component in ScalPy. As thawing models of scalar fields are more consistent with different observational data, we confine ourselves only to the thawing class of fields where the scalar field is initially frozen  due to large Hubble friction and starts very close to $w=-1$. ScalPy solves the autonomous system of equations for different kind of scalar field models with power-law and exponential potentials. It calculates various observables related to background expansion as well as to the growth of structures in universe. In this paper, we consider only the linear regime in the sub-horizon scales where effect of dark energy perturbation is negligible. Finally we integrate it with MCMC sampler ``emcee" to constrain various scalar field models with presently available data.

To conclude, we present a python package to study various scalar field dark energy models. This is probably the first package exclusively for the scalar field case. At present, we confine only to the canonical scalar fields and flat FRW models. But in near future, we shall incorporate various other scalar field models like coupled quintessence, scalar tensor models as well as non-flat  FRW models. We shall also include the observables related to super-horizon scales where dark energy perturbation can play an important role.

\begin{figure}
 {\includegraphics[scale=0.4]{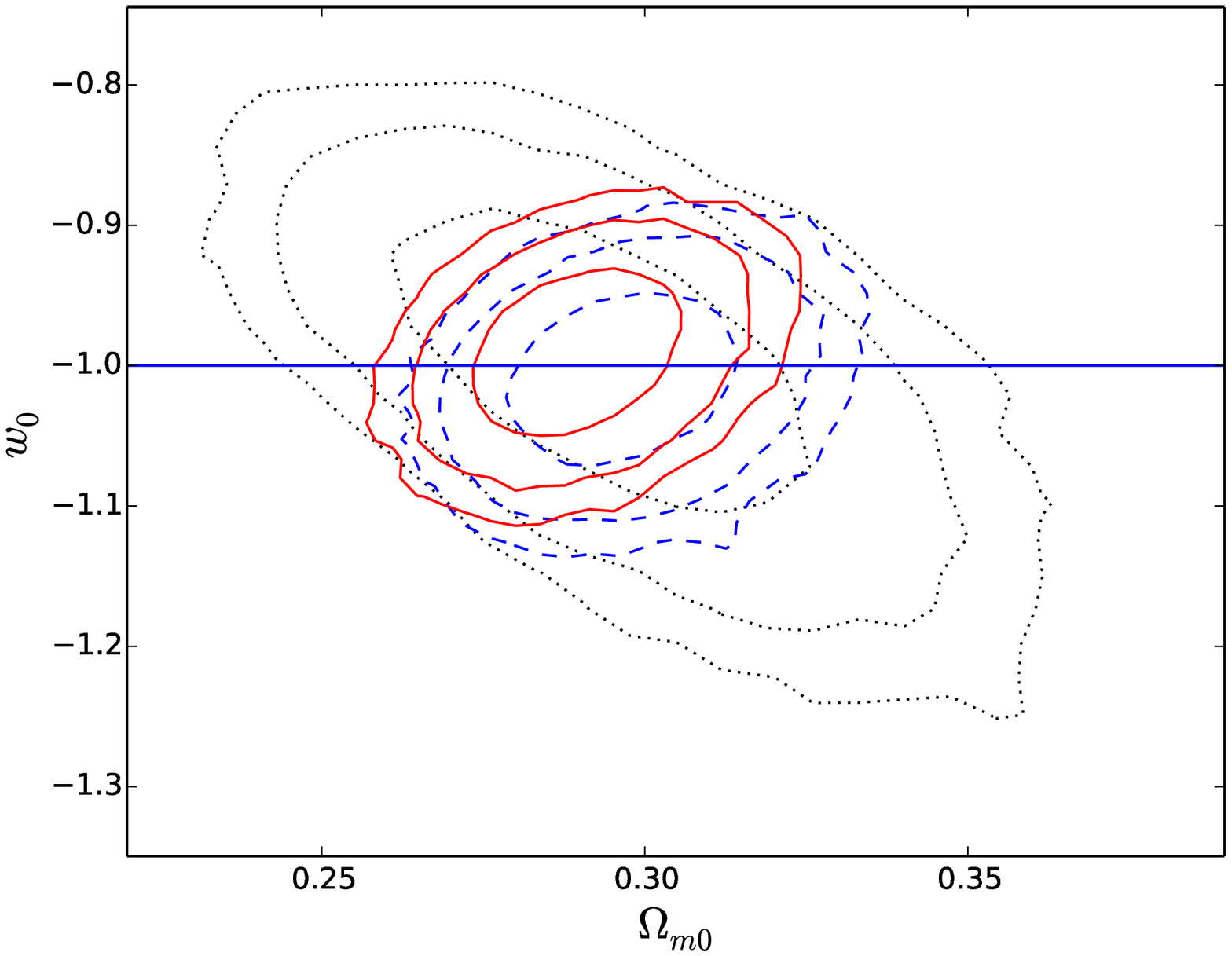}}
 \caption{1$\sigma$, 2$\sigma$ and 3$\sigma$ contours obtained for $JLA+BAO$(dotted),$JLA+BAO+CMB~Shift$ (dashed) and $JLA+BAO+CMB~Shift+f\sigma_8$ (solid) on model parameters for $w$CDM model.}
\end{figure}

\begin{figure}
 \begin{center}
 \begin{tabular}{cc}
  & \\
 {\includegraphics[width=2.6in,height=2.4in,angle=0]{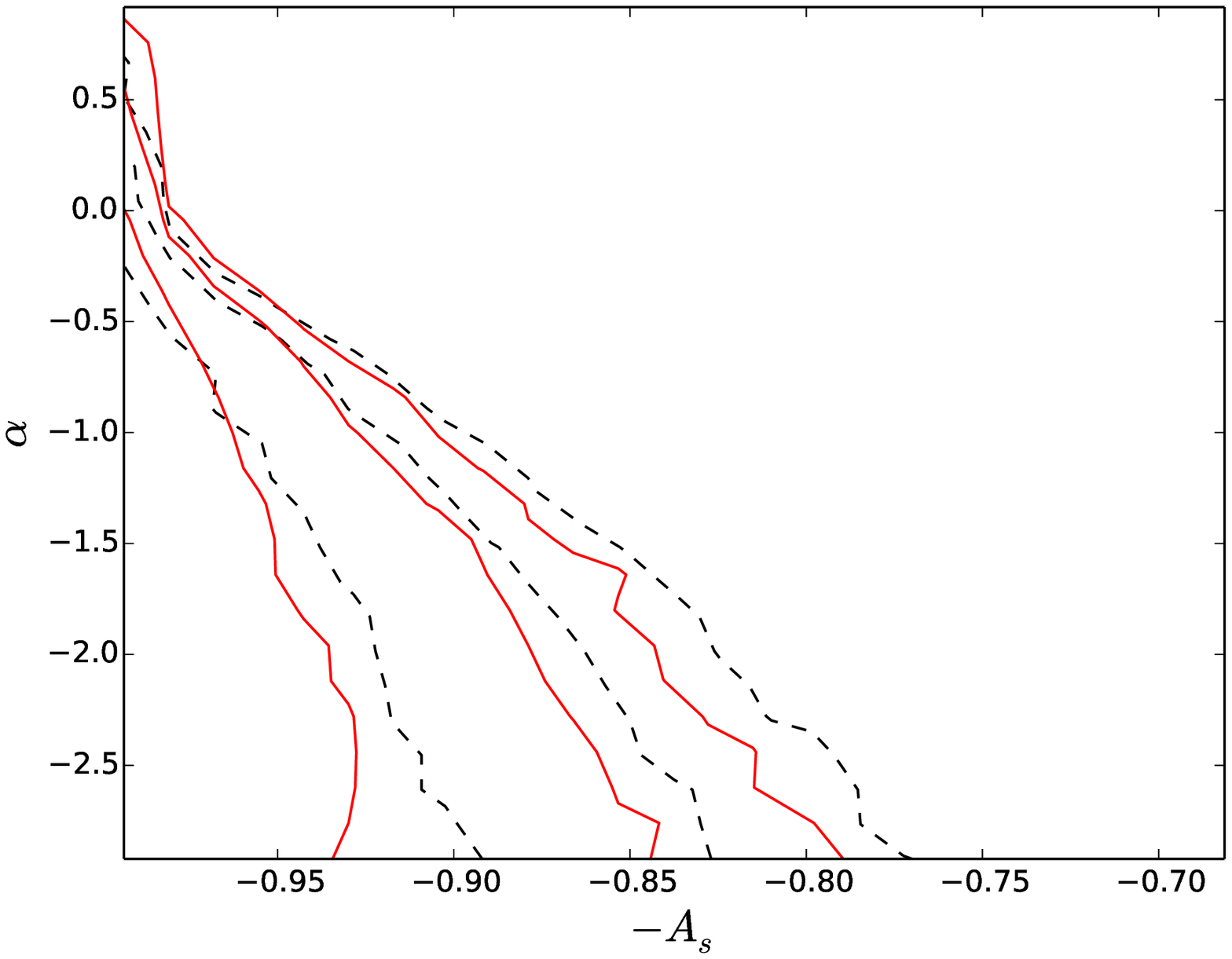}}&
 {\includegraphics[width=2.6in,height=2.4in,angle=0]{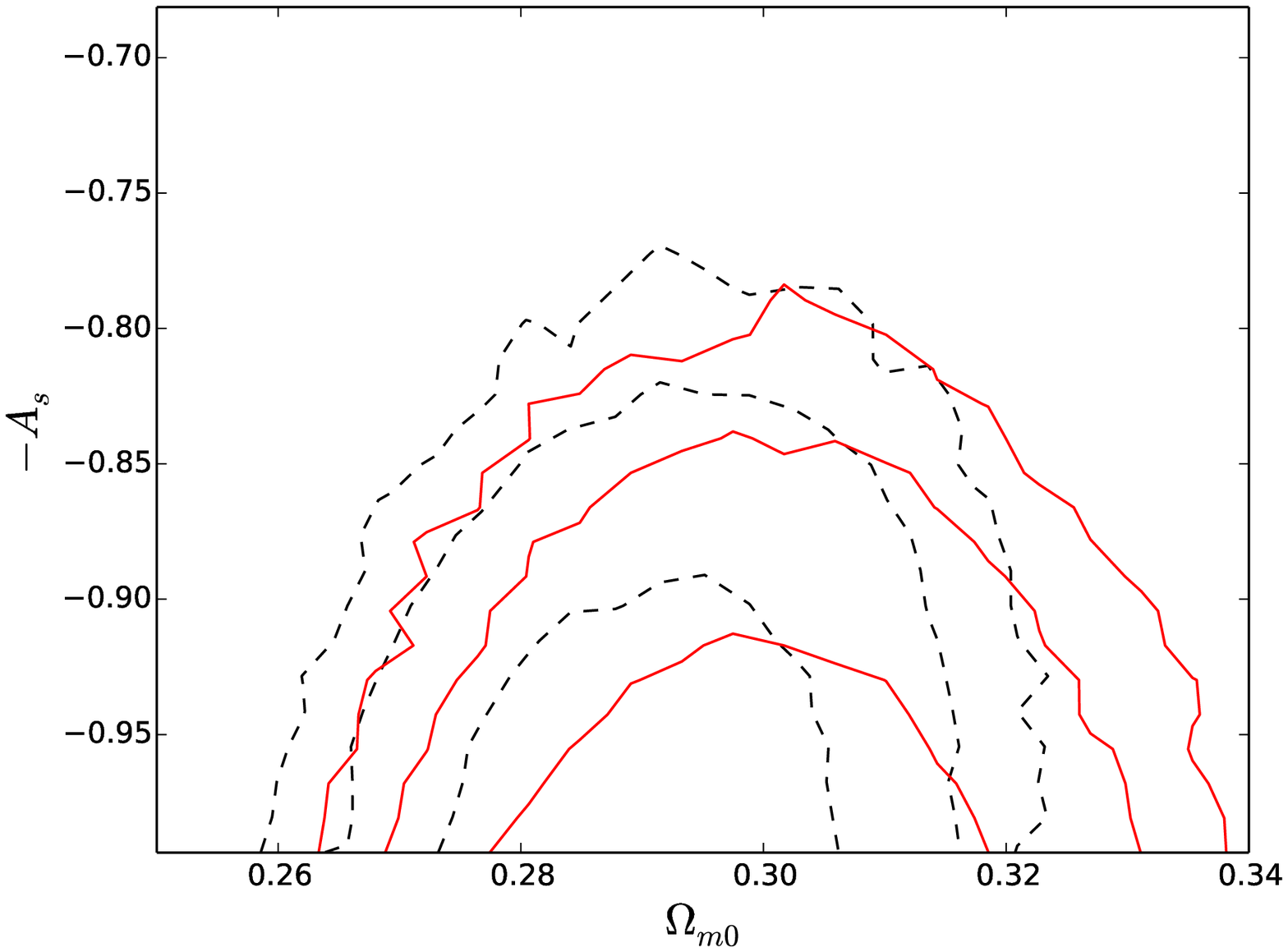}}
 \\

 \end{tabular}
 \caption{1$\sigma$, 2$\sigma$ and 3$\sigma$ contours obtained for $JLA+BAO+CMB~Shift$ (solid) and $JLA+BAO+CMB~Shift+f\sigma_8$ (dashed) on model parameters for GCG model. Present day equation of state $w_0$ is equal to $-A_s$ in GCG model.}
 \end{center}
 \end{figure}

\section{Acknowledgements}
AJ thank Centre for Theoretical Physics, Jamia Millia Islamia for kind hospitality where this work was initiated. AJ also acknowledges Indian Academy of Sciences for funding through Summer Research Fellowship Program. SK is funded by University Grants Commission, Govt. of India through SRF scheme. AJ and SK thank Md Wali Hossain for useful discussions.

\appendix
\section{Installation}
This package is written in python and it solves dynamical system for three types of minimally coupled scalar fields- quintessence, tachyon and Galileon with exponential and power law potential. Given the initial conditions, this module solves for cosmological observables such as Hubble parameter, luminosity distance, angular diameter distance, lookback time, growth function, growth rate, $f\sigma_8(z)$ and power spectrum (with Eisenstein-Hu transfer function as well as BBKS transfer function).
We have also defined all above cosmological observables for standard cosmological fluids such as $\Lambda$CDM, $w$CDM, $w_0w_a$CDM and generalized Chaplygin gas (GCG) \cite{gcg}. Table 1 describes different cosmological models we have used in the code.

\begin{table}[h]
\caption{Different parametrizations for dark energy models}
\begin{tabular}{|l|l|l|}
\hline
model & parameters &Hubble equation \\ \hline
$\Lambda$CDM     & $\Omega_{m0}$	& $\frac{H(z)}{H_0} = \sqrt{\Omega_{m0}(1+z)^3 + (1-\Omega_{m0})}$     \\ 
$w$CDM     & $\Omega_{m0}$, $w$ & $\frac{H(z)}{H_0} = \sqrt{\Omega_{m0}(1+z)^3 + (1-\Omega_{m0})(1+z)^{-3(1+w)}}$    \\ 
$w_0w_a$CDM     & $w_0$, $w_a$ and $\Omega_{m0}$ & $\frac{H(z)}{H_0} = \sqrt{\Omega_{m0}(1+z)^3 + (1-\Omega_{m0})(1+z)^{3(1+w_0+w_a)}\exp({-3w_az/(1+z)})}$   \\
GCG & $\Omega_{m0}$, $A_s$ and $\alpha$ & $\frac{H(z)}{H_0} = \sqrt{\Omega_{m0}(1+z)^3 + (1-\Omega_{m0})(A_s + (1-A_s)(1+z)^{3(1+\alpha)})^{1/(1+\alpha)}}$ \\ \hline
\end{tabular}
\end{table}
To install the code, one can simply use {\it pip} or download the full source code and run `setup.py' file. On a linux system where pip is installed, scalpy can be installed as\newline
user@computer \textgreater~sudo pip install scalpy

To install it from source code which is downloadable as .zip file from github repository, simply download the source from\newline
\url{https://github.com/sum33it/scalpy}

After extracting the file, move to the folder `scalpy-master' and run `setup.py' file\newline
user@computer \textgreater~ unzip scalpy-master.zip \newline
user@computer \textgreater~ cd scalpy-master \newline
user@computer \textgreater~ python setup.py install

\section{Structure}
The package consists of following main modules linked with each other: \newline
1){\bf scalar.py} This module solves the dynamical system for scalar fields such as minimally coupled quintessence, tachyon and Galileon. We have included two types of potential for each scalar field: power law $V(\phi)=K\phi^n$ and exponential $V(\phi) = \exp(\alpha\phi)$ where $n$,$\alpha$ and $K$ are real constants \\
2) {\bf fluids.py} This gives different observables for standard cosmological models such as $\Lambda CDM$, $wCDM$, $w_0w_aCDM$ and $GCG$ In this file also, different observables such as luminosity distance, Hubble parameter,  angular diameter distance, growth rate, growth function, power spectrum are defined\\
3) {\bf solver.py} For a given values of $w_0$ and $\Omega_{\phi 0}$, this module solves for the initial conditions needed at decoupling for scalar field models.\\
4) {\bf transfer{\_}func.py} In this module, we have defined transfer functions given by Eisenstein and Hu as well as by BBKS. \\

Detailed online documentation with examples can be found at following url:\\
\url{http://ctp-jamia.res.in/download/scalpy}

One can also read readme.rst file for quick Introduction provided in tarball downloaded from github repository.

\end{document}